\documentclass[aps,twocolumn,showpacs,floats]{revtex4}
\usepackage{graphicx}
\usepackage{dcolumn}
\usepackage{bm}
\usepackage{color}
\usepackage{amsmath}
\usepackage{amssymb}
\usepackage{hyperref}

\begin{document}

\title{Anderson localization in optical lattices with correlated disorder}

\author{E. Fratini and S. Pilati$^{1}$}
\affiliation{$^{1}$The Abdus Salam International Centre for Theoretical Physics, 34151 Trieste, Italy}

\begin{abstract}
We study the Anderson localization of atomic gases exposed to simple-cubic optical lattices with a superimposed disordered speckle pattern.
The two mobility edges in the first band and the corresponding critical filling factors are determined as a function of the disorder strength, ranging from vanishing disorder up to the critical disorder intensity where the two mobility edges merge and the whole band becomes localized.
Our theoretical analysis is based both on continuous-space models which take into account the details of the spatial correlation of the speckle pattern, and also on a simplified tight-binding model with an uncorrelated distribution of the on-site energies.
The mobility edges are computed via the analysis of the energy-level statistics, and we determine the universal value of the ratio between consecutive level spacings at the mobility edge.
We analyze the role of the spatial correlation of the disorder, and we also discuss a qualitative comparison with available experimental data for interacting atomic Fermi gases measured in the moderate interaction regime.
\end{abstract}

\maketitle


\section{Introduction}\label{Introduction}
\begin{figure}
\begin{center}
\includegraphics[width=1.0\columnwidth]{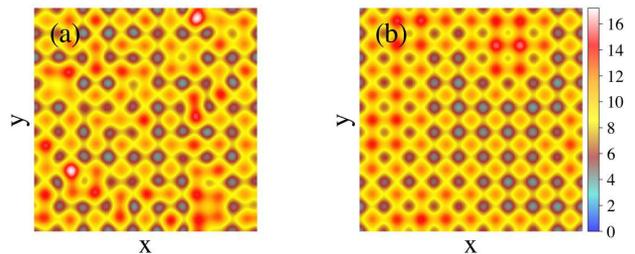}
\caption{(color online) 
Cross section of the three-dimensional intensity profiles of a  simple-cubic optical lattice with a superimposed blue-detuned isotropic optical speckle pattern. The optical lattice intensity is $V_0=4E_r$, the disorder strength is $V_{\mathrm{dis}}=1.3E_r$.  The speckles patterns in the two panels have different correlation lengths: $\sigma=d/\pi$ in panel (a) and $\sigma=d$ in panel (b). The color scale represents the potential intensity in units of recoil energy $E_r$.}
\label{fig1}
\end{center}
\end{figure}

Anderson localization, namely the complete suppression of wave diffusion due to sufficiently strong disorder~\cite{anderson1958absence}, is one of the most important and intriguing phenomena studied in condensed matter physics~\cite{lagendijk2009fifty,abrahams201050}.
Making reliable predictions for the critical disorder strength required to induce complete localization is a major theoretical challenge. 
In the theory of solid-state systems, studies that aim at a quantitative comparison between theory and experiments, and thus employ realistic models taking into account the details of a specific material, have appeared only recently~\cite{jarrellrealmaterials}.\\
Following the first experimental observations of Anderson localization in quantum matter waves~\cite{sanchez2007anderson,chabe2008experimental,billy2008direct,kondov2011three,jendrzejewski2012three}, ultracold atomic gases have emerged as the ideal setup to investigate the effects due to disorder in quantum systems~\cite{aspect2009anderson,sanchez2010disordered}. Feshbach resonances provide experimentalists with a knob to turn off the interatomic scattering, allowing them to disentangle the effects due to disorder from those due to interactions. Furthermore, using the optical speckle fields produced by shining coherent light through a diffusive plate, they can introduce disorder in a controlled manner, and even manipulate the structure of its spatial correlations~\cite{mcgehee2013three}; this kind of control is not possible in solid-state devices. Techniques to accurately measure the mobility edge, namely the energy threshold which separates the localized states from the extended states, have also been implemented~\cite{semeghini}.\\
Several previous theoretical studies on Anderson localization have disclosed the fundamental role played by the disorder correlations. In low dimensional systems, the characteristics of the correlations determine the presence or absence of an effective mobility edge~\cite{izrailev1999localization,piraud2013tailoring,lugan2009one,gurevich2009lyapunov,roemer,capuzzi2015enhancing}. In three dimensions, varying the correlation structure drastically changes the localization length and the transport properties~\cite{piraud2012matter,piraud2012anderson}.
In two recent studies, the mobility edge of ultracold atoms in the presence of isotropic and anisotropic optical speckle patterns has been precisely determined~\cite{delande2014mobility,fratini}, highlighting again the importance of taking into account the details of the disorder correlations. However, the experimental configuration which resembles more closely the behavior of electrons in solids is the one where the atoms are exposed to the deep periodic potential due to an optical lattice with,  additionally, the disorder due to a superimposed optical speckle pattern (see intensity profiles in Fig.~\ref{fig1}). 
This configuration with both an optical lattice and a speckle field has been implemented in experiments performed with Bose and Fermi gases~\cite{white2009strongly,pasienski2010disordered,kondov2015disorder}, so far considering interacting atoms.\\
%

\begin{figure}[h!]
\begin{center}
\includegraphics[width=1.0\columnwidth]{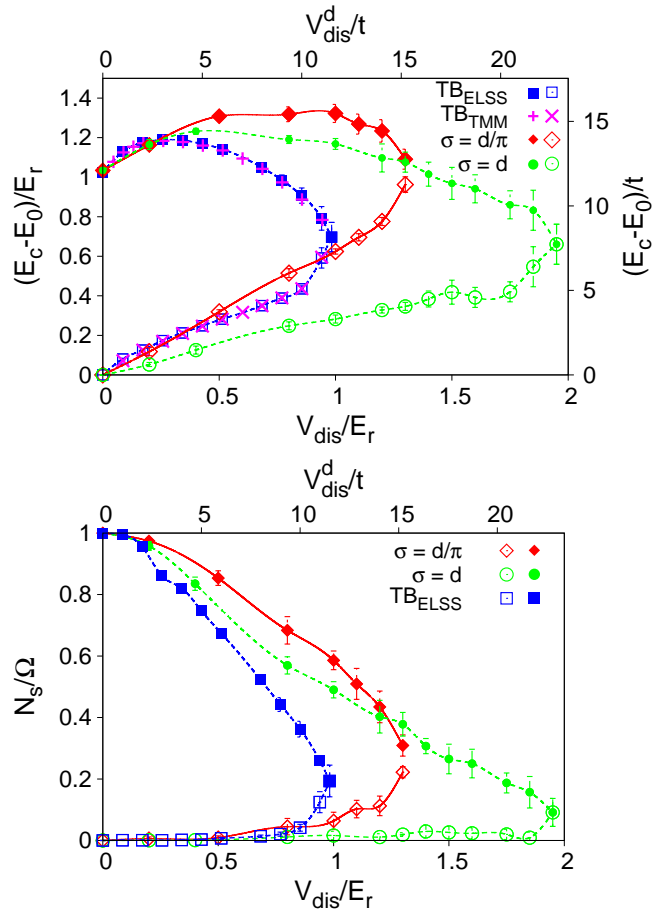}
\caption{(color online) 
Phase diagrams of an atomic gas exposed to three-dimensional simple-cubic optical lattices with a superimposed blue-detuned isotropic disordered speckle pattern:
(a) First two mobility edges $E_{c}$ as a function of the disorder strength $V_{\mathrm{dis}}/E_r$ (or $V_{\mathrm{dis}}^{\mathrm{d}}/t$ for the discrete-lattice model, in the top axis). 
Empty symbols correspond to the first mobility edge $E_{c1}$, full symbols to the second mobility edge $E_{c2}$ (see text). The energies are measured with respect to the bottom of the first band of the clean system $E_0$. 
The red rhombi and the green circles correspond to the continuos-space Hamiltonian~(\ref{h1})  with  correlation lengths $\sigma=d/\pi$ and $\sigma=d$, respectively. The blue squares correspond to our results for the tight-binding model with exponential on-site energy distribution, obtained via analysis of the energy-level spacings statistic ($\mathrm{TB_{ELSS}}$).  The results obtained in Ref.~\cite{pasek2015phase} using the transfer-matrix method ($\mathrm{TB_{TMM}}$) are represented by pink crosses. 
The optical lattice intensity is $V_0=4E_r$, and the corresponding hopping energy $t \cong  0.0855E_r$ is used to compare the continuous-space data (bottom-left axes) with the discrete-lattice data (top-right axes).
(b) Critical filling factors $N_s/\Omega$ as a function of the disorder strength, corresponding to the mobility edges represented in panel (a). $N_s$ is the number of states below the mobility edge, $\Omega$ the adimensional volume.
}
\label{fig2}
\end{center}
\end{figure}

%
In this Article, we investigate the Anderson localization of noninteracting atomic gases in a simple-cubic optical lattice plus an isotropic blue-detuned optical speckle field. The first two mobility edges and the corresponding critical filling factors are determined as a function of the disorder strength (see Fig.~\ref{fig2}). Our computational procedure is based on the analysis of the energy-level statistics familiar from quantum-chaos theory~\cite{haake2010quantum} and on the determination of the universal critical adjacent-gap ratio.\\
%
%
%
We employ both continuous-space models which describe the spatial correlation of an isotropic speckle pattern, and also an uncorrelated discrete-lattice model derived within a tight-binding scheme. This allows us to measure the important effect of changing the disorder correlations length, and to shed light on the inadequacy of the simple tight-binding approximation in the strong disorder regime.
Our (unbiased) results are important as a guide for future experiments performed with noninteracting atoms in disordered optical lattices, and also as a stringent benchmark for (inevitably approximate) theoretical calculations of the properties of disordered interacting fermions based on realistic models of disorder.\\
The rest of the Article is organized as follows:
in Section~\ref{Methods} we define our model Hamiltonians, describing the details of the optical speckle patterns; we explain our theoretical formalism and analyze the universality of the critical adjacent-gap ratio; furthermore, we provide benchmarks of our predictions with previous results for tight-binding models with box and with exponential disorder-intensity distributions.
In Section~\ref{Results} our predictions for the mobility edges and the critical filling factors are reported, with an analysis on the role played by the correlation length and on the validity of the tight-binding approximation. We also discuss the comparison with a recent transport experiment performed with atomic Fermi gases in the regime of moderate interaction strength~\cite{kondov2015disorder}.
Section~\ref{Conclusions} summarizes the main findings of this Article and reports our conclusions.
%

\section{Methods}\label{Methods}
We consider noninteracting atoms exposed to a simple-cubic optical lattice with a superimposed optical speckle pattern. 
The single-particle Hamiltonian which describes the system is:
 \begin{equation}
 \label{h1}
 \hat{H} = -\frac{\hbar^2}{2m}\Delta + V({\bf r}),
 \end{equation}
 where $\hbar$ is the reduced Planck's constant, $m$ is the atomic mass, and the external potential $V({\bf r}) = V_{\mathrm{L} }( \mathbf{r })  +  V_{\mathrm{S}} (\mathbf{r}) $ is the sum of the simple-cubic optical lattice $V_{\mathrm{L}} (\mathbf {r}=(x,y,z) )=V_0\sum_{\iota} \sin^{2}(\pi {\iota}/d)$ (here, $\iota=x,y,z$, $d$ is the lattice periodicity, and $V_0$ is the optical lattice intensity), and the disordered potential $V_{\mathrm{S}}(\bf{r})$ which represents the isotropic optical speckle pattern. 
 This intensity-based sum corresponds to the incoherent superposition of the optical-lattice and optical-speckle fields. In the following, it will be convenient to express $V_0$ in units of the recoil energy $E_r=\hbar^2/(2md^2)$. The size $L$ of the three-dimensional box is chosen to be a multiple of $d$, consistently with the use of periodic boundary conditions.\\
%
\begin{figure*}[t!]
\begin{center}
\includegraphics[width=\textwidth]{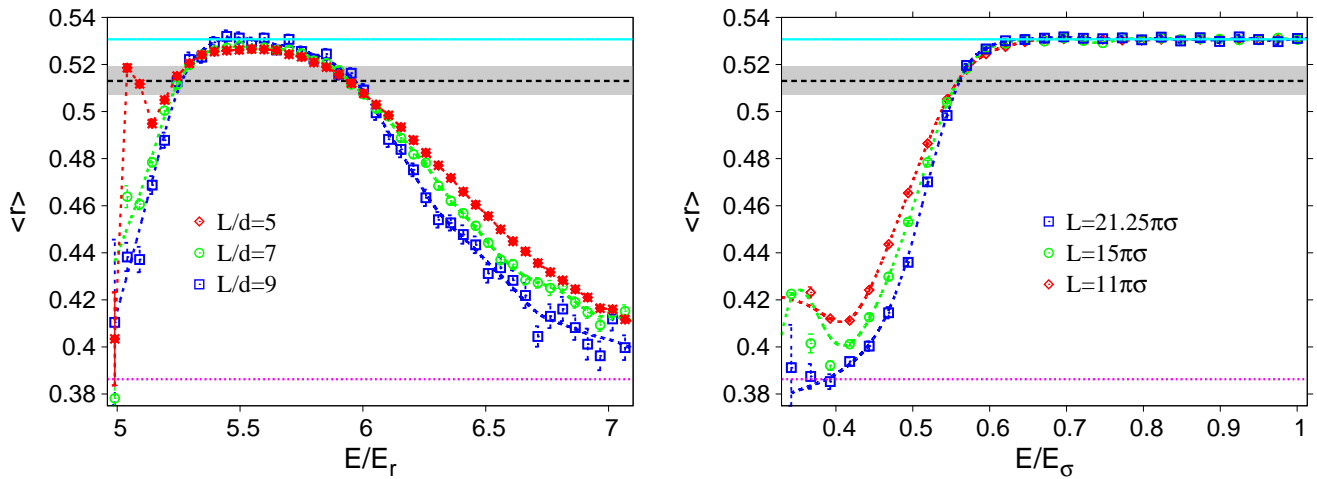}
\caption{(color online) 
Ensemble-average adjacent-gap ratio $\left<r\right>$ as a function of the energy $E$ for the continuous-space Hamiltonian~(\ref{h1}).
Left panel: simple-cubic optical lattice with intensity $V_0=4 E_r$ plus an isotropic optical speckle pattern with correlation length $\sigma=d/\pi$ and intensity $V_{\mathrm{dis}}=E_r$.
Right panel: optical speckle field with intensity $V_{\mathrm{dis}}=E_{\sigma}$ (without an optical lattice, namely $V_0=0$).
The three datasets correspond to different system sizes.
The horizontal cyan solid line indicates the value for the Wigner-Dyson distribution $\left<r\right>_{\textrm{WD}}$, the dashed magenta line the one for the Poisson distribution  $\left<r\right>_{\textrm{P}}$.
The dash-dot black line indicates the universal critical adjacent-gap ratio $\left<r\right>_C$, and the light-gray bar represents its error bar.
The energy units are the recoil energy $E_r$ and the correlation energy $E_{\sigma}$ (see text).
}
\label{fig3}
\end{center}
\end{figure*}

Disordered speckle patterns are realized in cold-atom experiments by shining lasers through diffusive plates, and then focusing the diffused light onto the atomic cloud~\cite{aspect2009anderson,sanchez2010disordered}. Fully developed speckle fields are characterized by an exponential distribution of the local intensities~\cite{goodman2007speckle}.
In the case of a blue-detuned optical field, the atoms experience a repulsive potential with the local-intensity distribution: $P_{\textrm{bd}}(V) = \exp\left(-V/V_{\mathrm{dis}}\right)/V_{\mathrm{dis}}$, if the local intensity is $V>0$, and $P_{\textrm{bd}}(V)=0$ otherwise. The (global) intensity parameter $V_{\mathrm{dis}}$ fixes both the spatial average of the disordered potential $V_{\mathrm{dis}}=\left<V_{\mathrm{S}}(\bf{r})\right>$ and also its standard deviation: $V_{\mathrm{dis}}^2=\left<V_{\mathrm{S}}({\bf r})^2\right>-\left<V_{\mathrm{S}}({\bf r})\right>^2$. For sufficiently large systems, spatial averages coincide with averages over disorder realizations.\\
The spatial correlations of the speckle pattern depend on the details of the illumination on the plate and of the optical setup used for focusing. We consider the idealized case of isotropic spatial correlations described by the following two-point correlation function~\cite{delande2014mobility}: $\Gamma(r=|{\bf r}|) = \left< V_{\mathrm{S}}({\bf r}'+{\bf r}) V_{\mathrm{S}}({\bf r}')\right>/V_{\mathrm{dis}}^2-1= \left[\sin(r/\sigma)/(r/\sigma)\right]^2$ (averaging over the position of the first point ${\bf r'}$ is assumed). The parameter $\sigma$ determines the length scale of the spatial correlations and, therefore, the typical size of the speckle grains. The full width at half maximum of the correlation function $\Gamma(r )$ (defined by the condition $\Gamma(\ell_c/2) =  \Gamma(0)/2$) is  $\ell_c \cong 0.89 \pi \sigma$, while the first zero is at $r_{\mathrm{z}}=\pi\sigma$. To generate this isotropic speckle pattern we employ the numerical recipe described in Ref.~\cite{fratini}. For further details on speckle pattern generation, see Refs.~\cite{huntley1989speckle,modugno2006collective,delande2014mobility}.\\
%
%
\begin{figure*}[t!]
\begin{center}
\includegraphics[width=\textwidth]{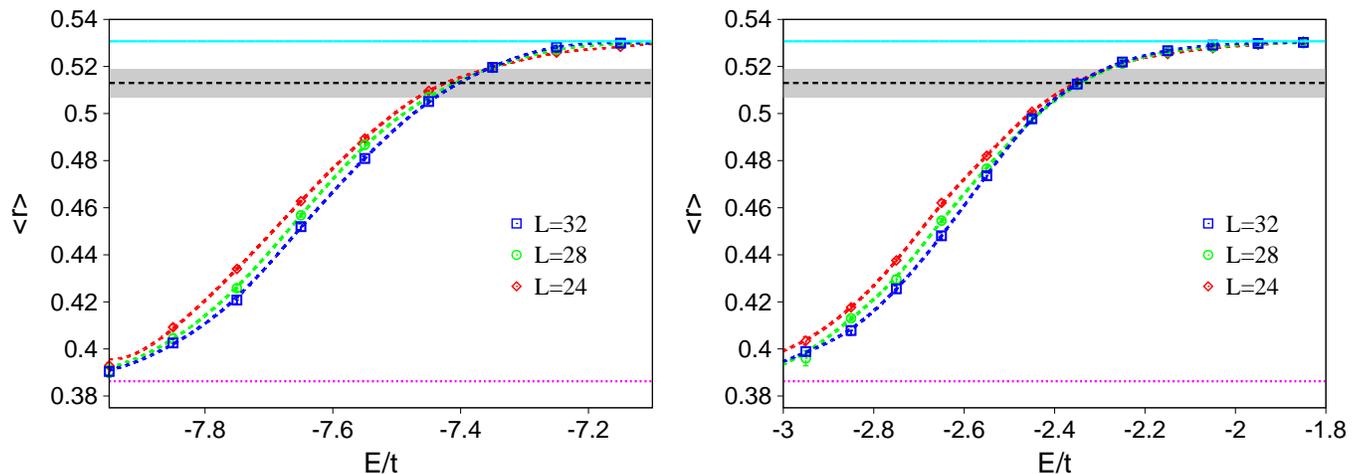}
\caption{(color online) 
Ensemble-average adjacent-gap ratio $\left<r\right>$ as a function of the energy $E$  for the tight-binding Hamiltonian~(\ref{h2}).
Left panel: three dimensional Anderson model with box disorder distribution with intensity $V_{\mathrm{dis}}^{\mathrm{d}} = 5t$.
Right panel:  three dimensional Anderson model with the exponential distribution with intensity $V_{\mathrm{dis}}^{\mathrm{d}} = 7t$. 
The three datasets correspond to different system sizes.
The horizontal cyan solid line indicates the value for the Wigner-Dyson distribution $\left<r\right>_{\mathrm{WD}}$, the dashed magenta line the one for the Poisson distribution  $\left<r\right>_{\mathrm{P}}$.
The dash-dot black line indicates the universal critical adjacent-gap ratio $\left<r\right>_C$, and the light-gray bar represents its error bar.
The energy unit is the hopping energy $t$.
}
\label{fig4}
\end{center}
\end{figure*}

We determine the positions of the mobility edges by analyzing the statistical distribution of the spacings between consecutive energy levels. The spectrum is obtained via exact diagonalization of the Hamiltonian matrix represented in momentum space, using the PLASMA library~\cite{plasma} for large-scale linear algebra computations on multi-core architectures. Special care is taken in analyzing the convergence of the results with the basis-set size. 
Further details on the numerical procedure can be found in Ref.~\cite{fratini}.\\
The mobility edges can be identified as the energy thresholds where the level-spacing distribution transforms from the Wigner-Dyson distribution characteristic of chaotic systems in the ergodic phase, to the Poisson distribution characteristic of localized systems, or vice versa~\cite{haake2010quantum}.
To distinguish the Wigner-Dyson and the Poisson distributions, it is convenient to consider the parameter $r= \min \left\{ \delta_n, \delta_{n-1} \right\}/\max \left\{ \delta_n, \delta_{n-1} \right\}$, where $\delta_n=E_{n+1}-E_{n}$ is the spacing between the $(n+1)$th and the $n$th energy levels, ordered for ascending energy values~\cite{oganesyan2007localization}. Its average over disorder realizations (later on referred to as adjacent-gap ratio) is known to be $\left< r \right>_{\mathrm{WD}} \simeq 0.5307$ for the Wigner-Dyson distribution and $\left< r \right>_{\mathrm{P}} \simeq 0.38629$  for the Poisson distribution~\cite{atas2013distribution}.\\
While in an infinite system $\left< r \right>$ would change abruptly at the mobility edge $E_c$, in finite systems one observes a smooth crossover from $\left< r \right>_{\mathrm{P}}$ to $\left< r \right>_{\mathrm{WD}}$, or vice versa. The critical point can be determined from the crossing of the curves representing $\left<r\right>$  versus energy $E$ corresponding to different system sizes $L$. We fit the data using the scaling function $\left< r \right> = g\left[ \left(E-E_c\right)L^{1/\nu}\right]$ (universal up to a rescaling of the argument)~\cite{abrahams201050} , where $\nu$ is the critical exponent of the correlation length. We Taylor expand the function $g[x]$ up to second order and obtain $E_c$ from the best-fit analysis.\\
This procedure to determine the mobility edges, which was previously employed in Ref.~\cite{fratini} for speckle patterns without optical lattices, requires several datasets corresponding to different system sizes with very small statistical error bars. A less computationally expensive procedure is obtained by exploiting the universal properties of the critical point. Indeed, the level-spacing distribution at the critical point differs both from the Wigner-Dyson and from the Poisson distributions~\cite{PhysRevB.47.11487,kravtsov1994universal}; it is expected to be system-size independent and universal, meaning that it does not depend on the details of the disorder. This implies a universal value of the critical adjacent-gap ratio, which we denote as $\left<r\right>_{\mathrm{C}}$, different from $\left< r \right>_{\mathrm{WD}}$ and from $\left< r \right>_{\mathrm{P}}$. 
We verified this universality by performing the finite-size scaling analysis for various models, determining $\left<r\right>_{\mathrm{C}}$ as the value of the scaling function at vanishing argument $g[0]$. In Fig.~\ref{fig3} we report the finite-size scaling analysis for a simple-cubic optical lattice with a superimposed speckle pattern, and also for a speckle pattern without the optical lattice (data from Ref.~\cite{fratini}). The critical adjacent-gap ratios $\left<r\right>_C$ of the two models (for the disordered optical lattice we consider the first two mobility edges) agree within statistical error bar. Furthermore, we verified that  $\left<r\right>_C$ does not depend on the disorder strength $V_{\mathrm{dis}}$, and that a compatible value of  $\left<r\right>_C$ is obtained also for red-detuned optical speckle fields, which have the same spatial correlations of blue-detuned speckle fields $\Gamma(r)$ defined above, but the opposite local-intensity distribution $P_{\mathrm{rd}}(V)=P_{\mathrm{bd}}(-V)$.\\

A further verification of the universality of the critical adjacent-gap ratio $\left<r\right>_C$ can be obtained by considering single-band models in a tight-binding scheme. The corresponding discrete-lattice Hamiltonian can be written in Dirac notation as:
\begin{equation}
 \label{h2}
\hat{H}_{\mathrm{d}} = -t \sum_{\left<i,j\right>}  \left| i \right> \left< j \right| +    \sum_{i} V_i \left| i \right> \left< i \right|  ,
\end{equation}
where the indices $i,j=1,\dots,L$ label the sites of the cubic discrete lattice of adimensional volume $\Omega=L^3$, $t$ is the hopping energy, and the brackets $\left<i,j\right>$ indicate nearest neighbor sites.  The on-site energies $V_i$ are chosen according to a random probability distribution. The most commonly adopted choice in the theory of Anderson localization is the box distribution $P_{\mathrm{b}}(V_i)=\theta(\left| V_i- V_{\mathrm{dis}}^{\mathrm{d}}\right|)/(2V_{\mathrm{dis}}^{\mathrm{d}})$. The parameter $V_{\mathrm{dis}}^{\mathrm{d}}$ determines the disorder strength. We also consider the exponential distribution $P_{\mathrm{e}}(V_i)=\exp\left(V_i/V_{\mathrm{dis}}^{\mathrm{d}}\right)/V_{\mathrm{dis}}^{\mathrm{d}}$, analogous to the exponential distribution $P_{\mathrm{bd}}(V)$ described above for blue-detuned speckle patterns in the continuous-space Hamiltonian. This discrete-lattice model with the exponential on-site energy distribution  is relevant to describe deep optical lattices with superimposed weak and uncorrelated speckle patterns, as explained more in detail in the Section~\ref{Results}.\\
The finite-size scaling analyses for these two lattice models (box and exponential distributions) are shown in Fig.~\ref{fig4}. The spectrum is obtained via exact diagonalization of the matrix representing the Hamiltonian $\hat{H}_{\mathrm{d}}$, defined on the three dimensional lattice. The universality of the critical adjacent-gap ratio is, again, confirmed within statistical uncertainty.\\

The average of the critical adjacent-gap ratios of the various models described above, including both the continuous-space models with correlated speckle patterns and the uncorrelated tight-binding models, is $\left<r\right>_C=0.513\pm0.05$; the error bar represents the standard deviation of the population. This prediction provides us with a computationally convenient criterion to locate the transition, consisting in identifying the mobility edge $E_c$ as the energy threshold at which the adjacent-gap ratio crosses the critical value $\left<r\right>_C$; the standard deviation of $\left<r\right>_C$ will be used to define the error bar on $E_c$. By applying this criterion to the isotropic speckle pattern (without optical lattice) analyzed in Fig.~\ref{fig3}, we obtain $E_c = 0.562(10)E_\sigma$ ($E_\sigma=\hbar^2/m\sigma^2$ is the correlation energy), in agreement with the transfer matrix theory of Ref.~\cite{delande2014mobility}, which predicts $E_c =0.570(7)E_\sigma$. We further confirm the validity of this criterion by reproducing the complete phase diagram of the discrete-lattice model with box disorder distribution (typically refereed to as Anderson model), making comparison with older results obtained using transfer-matrix theory~\cite{bulka1987localization} and multi-fractal analysis~\cite{grussbach1995determination}, as well as with the recent data from Ref.~\cite{ekuma2014typical} obtained using the typical medium dynamical cluster approximation; see Fig.~\ref{fig5}. Furthermore, in the case of the exponential disorder distribution, our results perfectly agree with the very recent transfer-matrix theory from Ref.~\cite{pasek2015phase} (see Fig.~\ref{fig2}).\\
It is worth specifying that our prediction for the universal critical adjacent-gap ratio $\left<r\right>_C$ applies to a cubic box with periodic boundary conditions. In fact, it has been predicted that the critical energy-level distribution, and so possibly the corresponding value of $\left<r\right>_C$, depends on the box shape~\cite{potempa1998dependence} and on the boundary conditions~\cite{PhysRevLett.81.1062,schweitzer1999influence}.

\section{Results}\label{Results}

\begin{figure}[t!]
\begin{center}
\includegraphics[width=1.0\columnwidth]{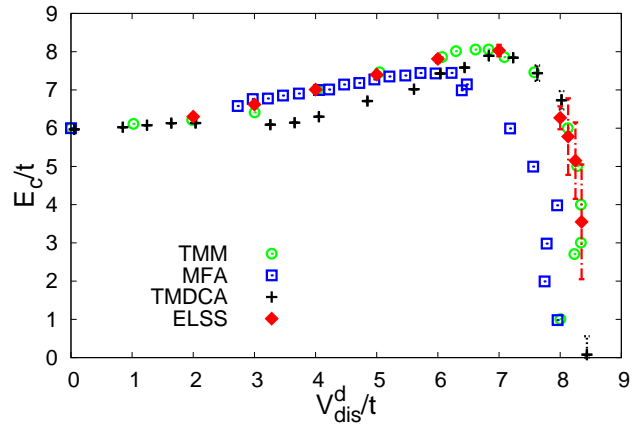}
\caption{(color online) 
Mobility edge $E_c$ as a function of the disorder strength $V_{\mathrm{dis}}^{\mathrm{d}}$ for the three dimensional Anderson model with box disorder distribution. 
Our data computed via the analysis of the energy-level spacings statistics  (ELSS, red diamonds) are compared with previous results obtained via transfer-matrix method (TMM, green circles, from Ref.~\cite{bulka1987localization}), via multi-fractal analysis (MFA, blue squares, from Ref.~\cite{grussbach1995determination}), and via typical medium dynamical cluster approximation (TMDCA, black crosses, from Ref.~\cite{ekuma2014typical}). The energy unit is the hopping energy $t$.}
\label{fig5}
\end{center}
\end{figure}

The continuous-space Hamiltonian~(\ref{h1}) accurately describes atomic gases exposed to optical lattices with superimposed optical speckle patterns, for any optical lattice intensity $V_0$ and disorder strength $V_\mathrm{dis}$. In particular, it takes into account the spatial correlations of the optical speckle pattern. In order to make comparison with recent experimental data, we consider the intermediate optical lattice intensity  $V_0=4E_r$, and we determine the lowest two mobility edges as a function of the disorder strength $V_\mathrm{dis}$, up to the critical value where the two mobility edges merge and the whole band becomes localized.\\

We consider two isotropic speckle patterns with correlation lengths $\sigma=d/\pi$ and $\sigma=d$. We recall that the first zero of the spatial correlation function $\Gamma(r)$ (see definition in Section~\ref{Methods}) is at $r_z = \sigma\pi$. Beyond this distance the speckle-field intensities are almost uncorrelated.
The intensity profiles of the total potential $V({\bf r})$ corresponding to these two correlation lengths are shown in Fig.~(\ref{fig1}).
The deformation of the regular structure of the simple-cubic optical lattice due to the speckle pattern is evident in both cases. In the first case the intensity values in nearest-neighbor wells of the optical lattice are only weakly correlated, while in the second case the correlations extend to a few lattice sites.\\
The phase diagram obtained using the methods presented in Section~\ref{Methods}, namely the analysis of the energy-level spacing statistics and the universality of the critical adjacent-gap ratio, is presented in Fig.~\ref{fig2}. The empty symbols indicate the lowest mobility edge $E_{c1}$, where the orbitals transform from localized (for energies $E<E_{c1}$) to extended (for $E>E_{c1}$), while the solid symbols indicate the second mobility edge $E_{c2}>E_{c1}$, where the orbitals transform from extended (for $E<E_{c2}$) to localized (for $E>E_{c2}$). Other mobility edges are located at significantly higher energies, outside the energy range investigated in this Article.\\ 
%
\begin{figure}[h!]
\begin{center}
\includegraphics[width=1.0\columnwidth]{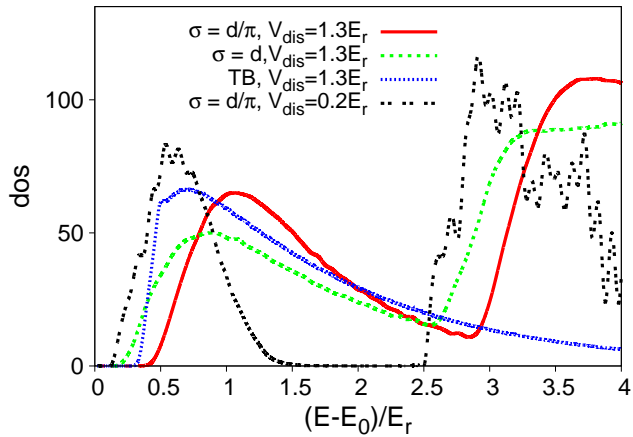}
\caption{(color online) 
Density of states $dos$ (in arbitrary units) as a function of the energy $E$ measured from the bottom of first band of the clean system $E_0$. The energy unit is the recoil energy $E_r$.
The  continuous red, dashed green, and the double-dash black curves correspond to the continuos-space model~(\ref{h1}) with different correlation lengths $\sigma$ and disorder intensities $V_{\mathrm{dis}}$. The dotted blue curve corresponds to the tight-binding (TB) model.}
\label{fig6}
\end{center}
\end{figure}

The data reported in Fig.~\ref{fig2} shed light on the fundamental role played by the spatial correlations of the disorder pattern. The critical disorder strength beyond which the first band is fully localized strongly depends on the correlation length. Indeed, for the short correlation length $\sigma=d/\pi$, full localization occurs already at the disorder strength $V_{\mathrm{dis}}\simeq 1.32 E_r$, while for the longer correlation length $\sigma=d$ full localization occurs only at the much stronger disorder intensity $V_{\mathrm{dis}}\simeq 1.95 E_r$.  This indicates that the disorder is more effective in inhibiting particle diffusion if the correlation length is short compared to the lattice spacing; also, it implies that, in order to quantitatively describe experiments performed with noninteracting atomic gases exposed to disordered optical lattices, it is necessary to take into account the details of the optical speckle pattern. 
%
%
In Fig.~\ref{fig2}, we report the two critical filling factors (defined as the number of eigenstates $N_s$ per adimensional volume $\Omega=(L/d)^3$ with energy $E<E_{c1}$ and $E<E_{c2}$) as a function of the disorder strength. The role of the spatial correlations is again manifest. Both the mobility-edge data and the critical filling-factors data display a strong asymmetry around the band center; this originates from the asymmetry of the exponential intensity distribution of the optical speckle pattern $P_{\mathrm{bd}}(V)$.\\

Most theoretical studies of atomic gases exposed to clean optical lattices are based on single-band tight-binding Hamiltonians analogous to the one defined in Eq.~(\ref{h2}).
The conventional procedure to map optical lattice systems to tight-binding models is based on the computation of the maximally localized Wannier function from the band-structure analysis of the periodic system. For sufficiently deep optical lattices $V_0\gg E_r$, the effect of higher Bloch bands and of hopping processes between non-adjacent Wannier orbitals can be ignored, leading to single-band tight-binding models in the discrete-lattice form defined by Eq.~(\ref{h2}). At the optical lattice intensity addressed in this Article, namely $V_0=4E_r$, the deep-lattice condition is marginally fulfilled, with a next-nearest neighbor hopping energy $\left| t_2\right| \cong 6.1\cdot10^{-3}E_r$, which is only one order of magnitude smaller than the nearest neighbor hopping energy $t \cong  0.0855E_r$.\\ 
In the presence of additional disordered optical fields, the conventional mapping procedure~\cite{jaksch1998cold,jaksch2005cold} based on band-structure calculation cannot be applied. A more generic approach, valid also in the presence of weak optical speckle patterns with intensity $V_{\mathrm{dis}}\ll V_0$, has been developed in Ref.~\cite{zhou2010construction}; this method allows one to construct an orthonormal basis of localized Wannier-like orbitals which describes the correct low-energy properties of weakly disordered optical-lattice systems. In the corresponding discrete-lattice Hamiltonian, the on-site energies $V_i$ have, with good approximation, the exponential distribution $P_{\mathrm{e}}(V_i)$, with a disorder intensity $V_{\mathrm{dis}}^{\mathrm{d}}\simeq V_\mathrm{dis}$, essentially coinciding with the intensity of the optical speckle field $V_\mathrm{dis}$. The on-site energies on nearby lattice sites have significative correlations which depends on the details of the optical speckle pattern. Also, the nearest-neighbor hopping energies have an (asymmetric) random distribution, characterized by strong correlations with the difference between the on-site energies of the corresponding lattice sites. In first approximation, one might neglect the hopping-energy fluctuations and the on-site energy correlations, and retain only the exponential on-site energy distribution. This approximate model of optical lattices with superimposed speckle patterns - which leads (in the noninteracting case) to the tight-binding Hamiltonian~(\ref{h2}) with the on-site energy distribution $P_{\mathrm{e}}(V_i)$ - has been adopted in Ref.~\cite{scarola2015transport} to describe a recent transport experiment performed with interacting ultracold atoms~\cite{scarola2015transport}. 
In this experiment, a drifting force was applied by introducing a magnetic-field gradient for a short interval of time; after this impulse, the confining potential was switched off, and the velocity of the center of mass of the atomic cloud was measured by absorption imaging and band mapping after a time of flight; the measurement was repeated with different intensities of the optical speckle field . Also, various optical lattice intensities were considered, ranging from $V_0=4E_r$ to $V_0=7E_r$.  The authors of Ref.~\cite{scarola2015transport} considered mainly the case of the deep optical lattice $V_0\simeq7E_r$, where the Hubbard interaction energy of two opposite-spin fermions on the same lattice site is large: $U\simeq 9t$. They argue that in this strongly interacting regime the details of the correlations of the hopping and of the on-site energies are not relevant, since transport is dominated by effective quasi-particles (not the original particles which are obviously relevant in the noninteracting case), which experience correlated hopping and interaction processes even in the simplified model. They indeed found satisfactory agreement between the computed center-of-mass velocities and the experimental data.\\
Our findings indicate that in the absence of interactions the details of the speckle pattern are, instead, important. The mobility edges of the uncorrelated tight-binding model~(\ref{h2}) (with the exponential on-site energy distribution) are shown in Fig.~\ref{fig2}, together with the results for the continuous-space model~(\ref{h1}). To make comparison between the two models, the energies in the lattice model have to be converted using the hopping energy $t\cong  0.0855E_r$ corresponding to the optical lattice intensity we consider, namely $V_0 = 4E_r$. One observes that certain qualitative features of the phase diagram are captured also by the tight-binding model. However, while at very weak disorder $V_{\mathrm{dis}}\approx 0.2E_r$ the continuous-space and the discrete-lattice models quantitatively agree, important discrepancies appear at strong disorder. In particular, the critical disorder strength where the whole band is localized in the discrete-lattice model, namely $V_\mathrm{dis}^d \simeq12t$ (corresponding to $V_\mathrm{dis} \simeq 0.95 E_r$), significantly underestimates  the results obtained with the more accurate correlated continuous-space models.\\
In principle, the details of the speckle pattern could be included also in a discrete-lattice Hamiltonian, following the numerical procedure of Ref.~\cite{zhou2010construction}. This approach has been adopted in Ref.~\cite{semmler2010localization} to investigate an interacting Anderson-Hubbard model with correlated speckle fields. However, the dynamical mean-field theory employed in Ref.~\cite{semmler2010localization} does not correctly describe the Anderson localization in the noninteracting limit, probably due to the assumed Bethe-lattice structure. More recently, the dynamical mean-field theory has been improved using the typical medium dynamical cluster approximation~\cite{ekuma2014typical}, allowing researchers to give more accurate predictions for the localization transition in the (uncorrelated) Anderson model with box distribution; the data from Ref.~\cite{ekuma2014typical} are reported in Fig.~\ref{fig5}.\\
Nevertheless, it should be emphasized that the numerical technique of Ref.~\cite{zhou2010construction} converges only as long as there is a well defined gap between the first and the second band. As shown in Fig.~\ref{fig6}, in our optical lattice the gap is well defined only for very weak disorder, while it is substantially filled when the intensity of the optical speckle field approaches the strength required to localize the whole band, making that numerical technique inapplicable.\\

\begin{figure}[h!]
\begin{center}
\includegraphics[width=1.0\columnwidth]{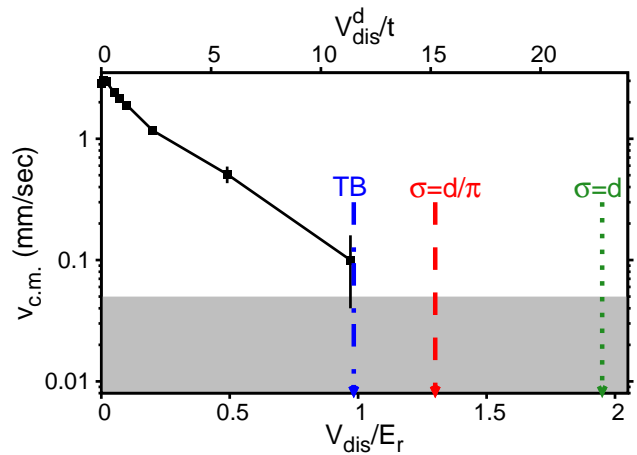}
\caption{(color online) 
Experimental data from Ref.~\cite{kondov2015disorder}: center of mass velocity $v_{\mathrm{c.m.}}$ of the atomic cloud (black squares) as a function of the disorder strength $V_{\mathrm{dis}}/E_r$ (or $V_{\mathrm{dis}}^{\mathrm{d}}/t$ for the tight-binding model, in the top axis). The vertical lines represent our predictions for the critical disorder strength where the whole band becomes localized in the continuos-space Hamiltonian~(\ref{h1}) (dashed red and dotted green lines) and in the uncorrelated tight-binding model~(\ref{h2}) with exponential disorder distribution (dot-dash blue line). The gray band represents the experimental resolution in detecting a vanishing velocity.}
\label{fig7}
\end{center}
\end{figure}

Experimental data for noninteracting atomic gases in disordered optical lattices are not available. However, in the experiment of Ref.~\cite{kondov2015disorder} (described above), which was performed with interacting atoms, the optical-lattice intensity was tuned down to $V_0=4E_r$, corresponding to a relatively small Hubbard interaction parameter, namely $U \simeq 2.3t$. It is then reasonable to discuss the comparison of these latter results with our theoretical predictions. It should be taken into account that the optical speckle pattern employed in the experiment is anisotropic, with an axial correlation length approximately $5$ times larger than the radial correlation length, and that its spatial correlations decays as a Gaussian function. However, the propagation axis of the optical speckle field is disaligned with respect to the optical lattice axes; this is expected to strongly reduce the role of the correlation anisotropy. If we consider the geometrically-averaged correlation length, we obtain a Gaussian correlation function with similar full width at half maximum as our speckle pattern with $\sigma=d$ (within $\approx15\%)$. Furthermore, in the experiment the density is inhomogeneous due to the confinement (with approximately $0.3-0.7$ particles per lattice well  in the trap center, per spin component) and the energy distribution is not precisely characterized.
In Fig.~\ref{fig7} we plot the center-of-mass velocities $v_{\mathrm{c.m.}}$ measured in the experiment, as a function of the disorder strength. 
The critical point where $v_{\mathrm{c.m.}}$ vanishes has been interpreted in Ref.~\cite{kondov2015disorder} as the average disorder strength required to localize the whole band, since all extended states are expected contribute to transport. We indeed observe that $v_{\mathrm{c.m.}}$ reaches negligible values (compatible with the experimental resolution) in the regime where we predict full localization to occur, depending on the details of the optical speckle pattern.
Clearly, a quantitative comparison with the experimental data would require a precise characterization of the experimental atomic density and of the energy distribution. This would also allow us to clarify the potential role played by states in higher-energy bands.
Nevertheless, this qualitative agreement between experimental data and theoretical predictions is encouraging, and should stimulate further experimental efforts aiming at observing Anderson localization in noninteracting atomic gases in disordered optical lattices. All details of the optical speckle pattern could be included in our theoretical formalism.

%
\section{Conclusions}\label{Conclusions}
In summary, we have investigated the Anderson localization of noninteracting atomic gases in disordered optical lattices. 
We considered both continuous-space models which describe the effect of a simple-cubic optical lattice with a superimposed isotropic blue-detuned optical speckle field, taking into account the spatial correlations of the disorder, and also an uncorrelated discrete-lattice Hamiltonian in a tight-binding scheme.\\
Our predictions for the mobility edges and for the critical filling factors indicate that the details of the speckle pattern play an important role; the critical disorder intensity where the whole band becomes localized strongly depends on the disorder correlation length. The tight-binding model with an uncorrelated (exponential) disorder distribution significantly underestimates this critical disorder strength.\\
Our theoretical formalism is based on the analysis of the energy-level statistics familiar from random-matrix and quantum-chaos theories and on the determination of the universal critical adjacent-gap ratio. The prediction for this universal value will be useful also in future studies of Anderson localization in different models belonging to the same universality class.\\
We have shown that the findings of a recent transport experiment performed with an atomic gas in the moderate interaction regime~\cite{kondov2015disorder} are qualitatively consistent with our predictions; this encouraging comparison should stimulate further experimental efforts to accurately measure the critical point of the Anderson transition in noninteracting atomic gases exposed to controlled and well characterized disordered fields. Such experiments would allow us to quantitatively benchmark sophisticated theories for Anderson localization based on realistic models which take into account all details of the disorder. This would be beneficial for the field of ultracold atoms, and likely beyond, possibly including the research on disordered materials, on randomized optical fibers~\cite{karbasi2012observation}, and on disordered photonic crystals~\cite{segev2013anderson}.\\

We acknowledge fruitful discussions with Brian DeMarco, Vito Scarola, Estelle Maeva Inack and Giuliano Orso. Brian DeMarco is also acknowledge for providing us the data from Ref.~\cite{kondov2015disorder}\\

\bibliographystyle{apsrev4-1}

\end{document}